\documentclass[aps,prb,groupedaddress,floatfix]{revtex4-2}

\usepackage[utf8]{inputenc}
\usepackage{bm}
\usepackage{amsfonts}
\usepackage{amsmath}
\usepackage{mathtools}
\usepackage{graphicx}
\usepackage{caption}
\usepackage{float}
\usepackage{subfiles}
\usepackage{color}

\begin{document}

\title{Computational Micromagnetics based on Normal Modes:\\  bridging 
the gap between macrospin and full spatial discretization}

\author{S. Perna$^1$,  F. Bruckner$^2$, C. Serpico$^1$, D. Suess$^2$ and M. d'Aquino$^1$}
\affiliation{1) Department of Electrical Engineering and ICT, University of Naples Federico II, Naples, Italy\\
2) University of Vienna Research Platform MMM Mathematics - Magnetism - Materials, University of Vienna,
Austria}

\begin{abstract}
The Landau-Lifshitz equation governing magnetization dynamics is written in terms of the amplitudes of normal modes associated with the micromagnetic system's appropriate ground state. 
This results in a  system of nonlinear ordinary differential equations (ODEs),
the right-hand side of which can be expressed as the sum of a linear
term and nonlinear terms with increasing order of nonlinearity (quadratic, cubic, etc.). The application of the method to nanostructured magnetic
systems demonstrates that the accurate description of magnetization dynamics requires a limited number of normal modes, which results in a considerable improvement in computational speed. 
The proposed method can be used to obtain a reduced-order dynamical description of magnetic nanostructures which allows to adjust the accuracy between low-dimensional models, such as macrospin, and micromagnetic models with full spatial discretization.
This new paradigm for micromagnetic simulations is tested for three problems relevant to the areas of spintronics and magnonics: directional spin-wave coupling in magnonic waveguides, high power ferromagnetic resonance in a magnetic nanodot, and injection-locking in spin-torque nano-oscillators. The case studies considered demonstrate the validity of the proposed approach 
to systematically obtain an intermediate order dynamical model based on normal modes for the analysis of magnetic nanosystems. The time-consuming calculation of the normal modes has to be done only one time for the system. These modes can be used to optimize and predict the system response for all possible time-varying external excitations (magnetic fields, spin currents).  This is of utmost importance for applications where fast and accurate system simulations are required, such as in electronic circuits including magnetic devices. 
\end{abstract}

\maketitle

\section*{Introduction}
Micromagnetism\cite{Brown_mumag},\cite{Aharoni},\cite{Kronmuller} is the most effective theory to predict 
magnetization behaviour of systems, at micrometer and sub-micrometer length scales, 
relevant to magnetic information and communication technologies\cite{Dieny2020}, 
and to the related research areas of spintronics\cite{Hirohata2020} and magnonics\cite{Magnonics2013}.  
A crucial role in the theory is played by the Landau-Lifshitz-Gilbert (LLG)  
equation\cite{Landau1935},\cite{Gilbert2004},\cite{BMS}, the nonlinear partial differential equation governing 
magnetization dynamics, and its generalizations\cite{Slonczewski1996},\cite{Zhang2004},\cite{Brataas2012}.
In most cases, the LLG equation is solved numerically due to nonlinearities and 
nontrivial  geometrical and physical  features of micromagnetic systems\cite{Schrefl2007},\cite{Abert2019}.
This has given rise to the area of Computational Micromagnetics\cite{Donahue1999,Chang2011, Andreas2014, Evans2014, Bisotti2018, Leliaert2019}.
There are special situations where the solutions of the LLG equation can be obtained analytically. 
This is usually possible when the geometrical dimensions of the magnet are not much larger than the exchange length, 
a characteristic length below which magnetization is  spatial uniform \cite{Brown1968},
and the LLG dynamics can be described in a low-dimensional state space.
This is the case of the macrospin model, in which magnetization is assumed
to be spatially uniform, used in the studying of magnetization switching\cite{StonerWohlfarth1948},\cite{Thirion2003}, 
ferromagnetic resonance\cite{Kittel1948},\cite{Sankey2006}, chaotic magnetization dynamics\cite{Montoya2019} and noise 
induced transitions\cite{Brown1963} in magnetic nanosystems. Other special cases are those related
to motion of domain walls or vortices which can be treated with the method of collective variables \cite{Tretiakov2008},\cite{Guslienko2010}. Recently, also neural networks have been used in order to predict magnetization dynamics of systems that are trained with full micromagnetic simulations \cite{Kovacs2019} or with single domain simulations \cite{Miao2019}. 
Except these notable cases, the solution of LLG is obtained by discretizing 
the body in a grid of small cells or finite elements with linear dimensions smaller than the exchange length.  
The LLG equation is then reduced to a nonlinear many-body evolution problem in which the state variables 
are the magnetization vectors defined on the grid.

This paper presents a new computational micromagnetic approach that bridges the gap between low-dimensional modeling, such as the Stoner-Wohlfarth model, and full spatially-discretized LLG equation. It relies on a spectral method where the degrees-of-freedom are not the magnetic moments associated with cells, but the amplitudes of {\em normal modes} obtained by the linearized LLG equation around the reference (ground) state of the problem under-investigation\cite{Grimsditch2004},\cite{McMichael2005},\cite{dAquino_JCP2009}. Although the expression of the LLG equation in terms of normal modes is mathematically equivalent to the original LLG, it turns out that in many situations of interest in the study of magnetic nanostructures, the confined nature of the problem leads to magnetization dynamics which can be described by a relatively small number of normal modes\cite{dAquino2009}. Thus, using the proposed spectral approach is possible to obtain a reduced dimensionality numerical model, which can be solved more efficiently than usual fully discretized micromagnetic problems. 
The normal modes-based method is also related to spin-wave theory, in which the deviation of magnetization field from the ground state is decomposed into a sum of Fourier-like plane-waves\cite{Herring1952}.
This approach is very convenient for the theoretical analysis of linear waves propagation in extended magnonic structures\cite{Stancil2009}. For confined structures, spin-wave theory is affected by
the difficulty that plane-waves do not correctly take into account boundary conditions\cite{Demokritov2007},\cite{Kalinikos}.
Normal modes, on the other hand, are obtained by solving the appropriate eigenvalues problem which includes
the correct boundary conditions\cite{dAquino_JCP2009}. In addition, the normal mode formulation of LLG enables to naturally take into 
account nonlinear coupling between the modes. In this respect,  the proposed approach allows extending the Suhl's analysis for the bulk magnet of nonlinear coupling of spin-waves\cite{Suhl1956,Suhl1957}
to the case of confined magnetic nanostructures.

We demonstrate the effectiveness of the proposed approach by studying nonlinear magnetization dynamics in three distinct magnetic structures which are relevant for applications in the emerging field of spintronics\cite{Dieny2020}, namely: directional spin-wave mode coupling in a magnonic waveguide, high-power ferromagnetic resonance (FMR) in a rectangular magnetic nanodot, and synchronization in injection-locked spin-torque oscillator.

The paper is organized as follows. First, the normal modes-based nonlinear model is introduced in connection with the Landau-Lifshitz-Gilbert equation with Slonczewski spin-torque\cite{Slonczewski1996} (LLGS). Then, it is shown that this model is able to make quantitative predictions on the aforementioned three magnetic systems using a small number of normal modes. These predictions are in very good agreement with full micromagnetic simulations. Finally, the advantages of the proposed model with respect to full micromagnetic simulations are  discussed.

\begin{table*}[t]
    \centering
    \begin{tabular}{|c|c|}\hline
         Landau-Lifshitz-Gilbert-Slonczewski (LLGS) equation (in dimensionless form) \\ \\
         
         $\frac{\partial \bm m}{\partial t} -\alpha_\mathrm{G}\bm m\times \frac{\partial\bm m}{\partial t}= -\bm m\times\bm h_\mathrm{eff}[\bm m]-\bm m\times\bm h_\mathrm{rf} +\beta\bm m\times(\bm m\times\bm p)$ \\
         \text{Effective field } $\bm h_\mathrm{eff}[\bm m]=-\mathcal{C}\bm m + \bm h_\mathrm{a} \quad,\quad \mathcal{C}=-l_\mathrm{ex}^2\nabla^2 + \mathcal{N}+\kappa_\mathrm{an}\bm e_\mathrm{an}\otimes \bm e_\mathrm{an}$ (exchange, magnetostatic, anisotropy),\\
         \text{Micromagnetic equilibrium (ground state)} $\bm m_0\times\bm  h_\mathrm{eff}[\bm m_0]=0, \text{ in } \Omega \quad,\quad \frac{\partial \bm m_0}{\partial \bm n}=0 \text{ on } \partial\Omega $,\\ 
         Radio-frequency applied field $\bm h_\mathrm{rf}(t)$ \,,\, spin-torque intensity $\beta$ with polarizer unit-vector $\bm p$. \\  \\ 
         
         \hline
          Normal Modes Model (NMM)  \\ \\
          Generalized eigenvalue problem\cite{dAquino_JCP2009}  $\mathcal{A}_{0\perp}\bm\varphi_k = \omega_k  \mathcal{B}_{0\perp}\bm\varphi_k \quad;\quad \bm\varphi_k,\omega_k$ eigenmodes, eigenfrequencies, \\ $\mathcal{A}_{0\perp}=\mathcal{P}_{m_0}(\mathcal{C} + h_0 \mathcal{I}) \,,\, \mathcal{P}_{\bm m_0} = \mathcal{I} - \bm m_0\otimes\bm m_0\,,\,\mathcal{B}_{0\perp}=-j\Lambda(\bm m_0) \,,\, h_0=\bm m_0\cdot\bm h_\mathrm{eff}[\bm m_0] \,,\, \Lambda(\bm m)\cdot\bm h=\bm m\times \bm h \,.$\\ 
          Orthogonality of Eigenmodes: $(\bm \varphi_k,\mathcal{A}_{0\perp}\bm \varphi_h) = \frac{1}{V_\Omega}\int_\Omega\bm \varphi_k^*\cdot,\mathcal{A}_{0\perp}\bm \varphi_h)\,dV=\delta_{kh}$\\
          Magnetization field decomposition: $\bm m(\bm r,t) = \bm m_0\left(1-\frac{1}{2}\delta\bm m_\perp^2\right) + \delta \bm m_\perp$, $\delta\bm m_\perp(\bm r,t) = \sum_k a_k(t)\bm\varphi_k(\bm r)$\\
          
          \\
Nonlinear mode dynamics ($a_h$ amplitude of $h-$th mode) equation \eqref{eq:norm_modes_model}\\         
               $\begin{aligned}
(1+\alpha_\mathrm{G}^2)\,\dot{a}_h = &j\omega_h\, b_{h} + j\omega_h\sum_i b_{hi}a_i + j\omega_h\sum_{i,j}c_{hij} a_ia_j\,+ \frac{j\omega_h}{2}\sum_{i,j,k}d_{hijk}a_ia_ja_k\,+\\
&\frac{j\omega_h}{4}\sum_{i,j,k,l}e_{hijkl}a_ia_ja_ka_l\,+ \frac{j\omega_h}{4}\sum_{i,j,k,l,m}f_{hijklm}a_ia_ja_ka_la_m\,,
\end{aligned}$\\ 

\\

 Coefficients $b,c,d,e,f$ (computed once for the system under investigation, see section Methods): \\ 
 
 $b_h=-(\bm\varphi_h,\bm f_c)+\,(\bm\varphi_h,\bm m_0\times\bm f_d)\,$,\\ 
$\bm f_c = \bm h_\mathrm{rf} + \alpha_\mathrm{G}\beta \bm p\,,\,\bm f_d = \alpha_\mathrm{G}\bm h_\mathrm{rf} -\beta\,\bm p$\\\\ 
 $b_{hi} = \delta_{hi}+\alpha_\mathrm{G} b_{hi}^\alpha + b_{hi}^f\,$,\\
  $b_{hi}^\alpha = j\,\omega_i\,\left(\bm\varphi_h, \bm\varphi_i\right)\,,$\hspace{1cm}
 $b_{hi}^f= -\left(\bm\varphi_h,\bm m_0\cdot\bm f_c\,\bm\varphi_i\right)-\left(\bm m_0\times\bm\varphi_h,\bm m_0\cdot \bm f_d\,\bm\varphi_i\right)$ \\\\ 
 
 $c_{hij}=c_{hij}^0 + \alpha_\mathrm{G} c_{hij}^\alpha +c_{hij}^f\,$,\\
 $c_{hij}^0 = -\frac{1}{2}\left(\bm\varphi_h,\mathcal{C}\psi_{ij}\bm m_0\right) -\left(\bm\varphi_h,\bm m_0\cdot\mathcal{C}\bm\varphi_j\bm\varphi_i\right)\,,\,\,\psi_{ij} = \bm\varphi_i\cdot\bm\varphi_j $\\
 $c_{hij}^\alpha = \frac{1}{2}\left(\bm m_0\times\bm\varphi_h,\mathcal{C}\psi_{ij}\bm m_0\right) +\left(\bm m_0\times\bm\varphi_h,\bm m_0\cdot\mathcal{C}\bm\varphi_j\bm\varphi_i\right)$\\
 $c_{hij}^f = \frac{1}{2}\left(\bm\varphi_h,\psi_{ij}\,\bm f_c\right) - \left(\bm m_0 \times\bm\varphi_h,\bm \varphi_j\cdot\bm f_d\,\bm\varphi_i\right)$\\\\
 
 $d_{hijk}=d_{hijk}^0 + \alpha_\mathrm{G} d_{hijk}^\alpha +d_{hijk}^f\,$\\
 $d_{hijk}^0 =-\left(\bm\varphi_h, \psi_{jk}\,\mathcal{C}\bm\varphi_i\right) + \left(\bm\varphi_h,\bm m_0\cdot\mathcal{C}\psi_{jk}\bm m_0\,\bm \varphi_i \right)$\\
 $d_{hijk}^\alpha = \left(\bm m_0\times\bm\varphi_h,h_0\, \psi_{jk}\,\bm\varphi_i\right)-\left(\bm m_0\times\bm\varphi_h,\bm m_0\cdot\mathcal{C}\psi_{jk}\bm m_0\bm\varphi_i\right) +2\left(\bm m_0\times\bm\varphi_h,\,\bm \varphi_k\cdot\mathcal{C}\bm\varphi_j\,\bm\varphi_i\right)$\\
 $d_{hijk}^f =\left(\bm m_0\times\bm\varphi_h,\psi_{jk}\,\bm m_0\cdot \bm f_d\,\bm\varphi_i\right)$
 \\\\ 
 $e_{hijkl} = e_{hijkl}^0 +\alpha_\mathrm{G} e_{hijkl}^\alpha\,$,\\
 $e_{hijkl}^0  =\left(\bm\varphi_h, \psi_{kl}\,\mathcal{C}\psi_{ij}\bm m_0\right) $\\
 $e_{hijkl}^\alpha = -2\left(\bm m_0\times\bm\varphi_h,\psi_{kl}\,\bm m_0\cdot\mathcal{C}\bm\varphi_j\bm\varphi_i\right) -2 \left(\bm m_0\times\bm\varphi_h,\bm \varphi_l\cdot\mathcal{C}\psi_{jk}\bm m_0\,\bm\varphi_i\right)$\\\\ 
 $f_{hijklm} = \alpha_\mathrm{G}f_{hijklm}^\alpha\,,$\\
 $f_{hijklm}^\alpha=\left(\bm m_0\times\bm\varphi_h,\psi_{lm}\,\bm m_0\cdot\mathcal{C}\psi_{jk}\,\bm m_0\bm\varphi_i\right)$\\
 \hline

    \end{tabular}
    \caption{Normal Modes-based Model in connection with LLGS equation. All the quantities are in dimensionless units. Derivation of parameters and coefficients can be found in the Methods section.}
    \label{tab:NMM}
\end{table*}

\section{Results}

The main result of this is work is the quantitative description of nonlinear magnetization dynamics in micromagnetic systems evolving around a given arbitrary (spatially inhomogeneous) ground state  by using reduced-order models with incremental accuracy controlled by the number of involved degrees-of-freedom, whose role is played by selected magnetization normal modes \cite{dAquino_JCP2009}. From the mathematical point of view, such description is formalized with a system of nonlinear differential equations where the unknowns are complex functions of time which are the amplitudes of the aforementioned selected normal modes. The detailed derivation of such a description is reported in the section Methods. In the following, we outline the main points instrumental to the illustration of the results. The starting point is the LLGS equation in dimensionless form (see table \ref{tab:NMM}), which models the dynamics of the magnetization (unit-vector) field $\bm m(\bm r,t)$ (normalized by the saturation magnetization) in the magnetic body $\Omega$ under investigation. Consistently with the fundamental micromagnetic constraint ($\bm m^2=1$), the magnetization field is decomposed according to the following equations (see section Methods for details):
\begin{equation}\label{eq:magn_exp}
\bm m(\bm r,t) = \delta \bm m_\perp(\bm r,t) + \bm m_0(\bm r)\left(1-\frac{1}{2}\delta\bm m_\perp^2-\frac{3}{4!}\delta\bm m_\perp^4+\dots\right)\,,
\end{equation}
where $\bm m_0(\bm r)$ is the (inhomogeneous) magnetic ground state (see table \ref{tab:NMM}), and $\delta\bm m_\perp(\bm r,t)$ is the projection of the magnetization on the plane pointwise perpendicular to the micromagnetic equilibrium $\bm m_0$. We refer to the class of vector fields perpendicular pointwise to $\bm m_0$ as $\mathcal{TM}(\bm m_0)$. The transverse magnetization $\delta\bm m_\perp(\bm r,t)$ can be expressed as:
\begin{equation}\label{eq:deltam_modes}
    \delta\bm m_\perp(\bm r,t)=\sum_h a_h(t) \bm\varphi_h(\bm r) \quad,
\end{equation}
where the set $\{\bm\varphi_1,\,\bm\varphi_2,\dots,\,\bm\varphi_h,\dots\}$ spans square-integrable vector fields defined in $\mathcal{TM}(\bm m_0)$ and the vectors $\bm\varphi_h$ are solutions of suitable generalized eigenvalue problem\cite{dAquino_JCP2009} (see table \ref{tab:NMM} and section Methods). 
From the knowledge of eigenmodes $\bm \varphi_h(\bm r)$ and time evolution of coefficients $a_h(t)$ (projection of $\delta\bm m_\perp(\bm r,t)$ on the generic eigenmode $\bm\varphi_h(\bm r)$) one can reconstruct the magnetization field 
by using eq.\eqref{eq:magn_exp}. 
It is worth  to remark that the eigenmodes can be determined once the ground state is defined, independently from the micromagnetic dynamics. In this framework, the complex amplitudes $a_h(t)$ are the degrees of freedom of the magnetization dynamics and their evolution is mathematically formulated by the following system of coupled nonlinear ordinary differential equations, here and on referred to as Normal Modes Model (NMM):
\begin{equation}\label{eq:norm_modes_model}
\begin{aligned}
\,\dot{a}_h = \frac{j\omega_h}{1+\alpha_\mathrm{G}^2}\, &\left(b_{h} + \sum_i b_{hi}a_i + \sum_{i,j}c_{hij} a_ia_j\,+ \frac{1}{2}\sum_{i,j,k}d_{hijk}a_ia_ja_k\,\right.+\\
&\left.\frac{1}{4}\sum_{i,j,k,l}e_{hijkl}a_ia_ja_ka_l\,+ \frac{1}{4}\sum_{i,j,k,l,m}f_{hijklm}a_ia_ja_ka_la_m + \dots\right)\,.
\end{aligned}
\end{equation}
The NMM can be derived by substituting eq.\eqref{eq:magn_exp} in the LLGS, then expressing the field $\delta\bm m_\perp(\bm r,t)$ as a function of eigenmodes (see eq.\eqref{eq:deltam_modes}) and complex amplitudes and finally using the orthogonality property of eigenmodes to project the LLGS equation on each eigenmode (see table \ref{tab:NMM}). In section Methods, such procedure is detailed for the case where the expansion \eqref{eq:magn_exp} is truncated at second-order terms in $\delta\bm m_\perp(\bm r,t)$. Such approximation has been termed parabolic approximation and the resulting NMM with all the coefficient expressions is also listed in table \ref{tab:NMM}. The highest order nonlinear terms involve the product of 5 complex amplitudes. In situations where magnetization goes far from the equilibrium beyond the validity of the parabolic approximation, terms with the product of a larger number of complex amplitudes may be added to the NMM in order to obtain more accurate description.\\
Differently from what happens for micromagnetic simulations where the number of the degrees of freedom reflects that of elementary magnetized cells, the NMM involves a preliminary computation of magnetization eigenpairs $(\omega_h,\bm \varphi_h)$ by solving suitable generalized eigenvalue problem (see table \ref{tab:NMM}, or eq.\eqref{eq:gen_eig_prob} in Methods) which depends exclusively on material and geometrical parameters and is performed only once for a given system under investigation.
The special nature of the generalized eigenvalue problem allows the determination of a prescribed number of eigenmodes\cite{dAquino_JCP2009} using Arnoldi-like iterative methods using classical computational routines adopted within micromagnetic codes in a large-scale fashion.
It is worthwhile to remark that this preliminary computation is independent from external forcing terms in LLGS equation (e.g. microwave field, spin currents, etc.). By using eigenpairs $(\omega_h,\bm \varphi_h)$ along with computational routines for computing the micromagnetic effective field, all the coefficients appearing in eq.\eqref{eq:norm_modes_model} can be also efficiently computed in the set-up phase.
\begin{figure*}
    \centering
    \includegraphics[width = 18 cm,height = 5 cm]{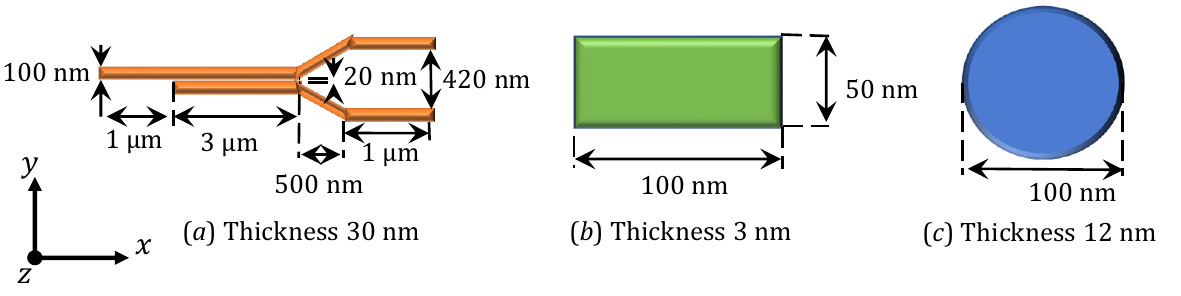}
    \caption{Sketch of the micromagnetic systems under investigation with the NMM: a) magnonic waveguide, b) rectangular nanodot and c) spintronic oscillator. }
    \label{fig:geom_3sys}
\end{figure*}
In the following, we show that eq.\eqref{eq:norm_modes_model} considering a limited number of normal modes and terms is able to reproduce with quantitative accuracy the results of full micromagnetic simulations. Moreover, we show that this description allows the development of general analytical formulas for magnetization dynamics which can be instrumental in optimization and design of magnetic technologies. 
In the sequel, the NMM is applied to the study of three model-problems in nonlinear magnetization dynamics which span relevant applications in the world of spintronics: nonlinear spin-wave coupling in a magnonic waveguide,  high-power ferromagnetic resonance in a rectangular magnetic nanodot, and synchronization in injection-locked spin-torque oscillator. The sketches of these devices with their dimensions are shown in fig.\ref{fig:geom_3sys}. In order to check the accuracy of the method, the results obtained by NMM are compared with full-scale micromagnetic simulations of the investigated systems.


\subsection{Directional spin-wave mode coupling in Magnonic Waveguide}\label{sec:magnonic waveguide}
\begin{figure*}
    \centering
    \includegraphics[width = 18 cm,height = 13 cm]{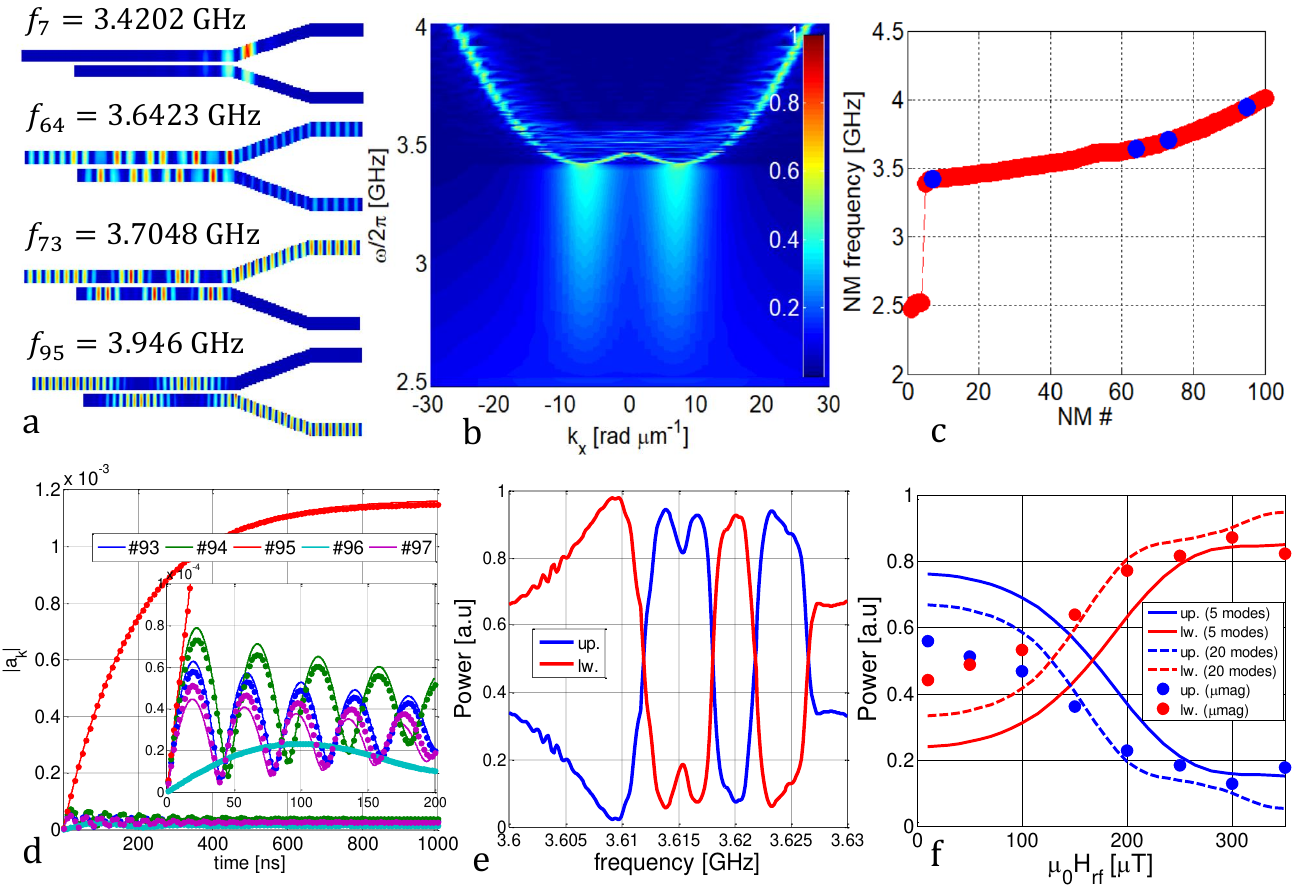}
    \caption{Normal mode analysis for the magnonic waveguide shown in fig.\ref{fig:geom_3sys}-(a). (a) Plots of the function $\|\bm \varphi\|^2$, normalized in the range $(0,1)$, for different spatial distributions of normal modes. In order from the top: confined mode, double arm mode, upper arm mode and lower arm mode respectively. The color scale is the same as for (b). (b) Dispersion curve. (c) Normal modes frequencies resulting from the solution of the eigenvalue problem \eqref{eq:gen_eig_prob}. Blue filled dots are the frequencies values corresponding to the normal modes plotted in (a). (d) Comparison between the prediction of equation \eqref{eq:lin_norm_modes_evol} (solid lines) and micromagnetic simulations (dots) for linear magnetization dynamics. (e) Power transmitted by the waveguide in the linear regime as a function of the excitation frequency. The red line indicates the power transmitted on the lower arm while the blue line the power transmitted on the upper arm. (f) Comparison of the power transmitted by the waveguide as a function of the microwave field amplitude, computed with the NMM and by micromagnetic simulations. The colors are coherent to (e). Solid and dashed lines refer to NMM integrated with 5 and 20 modes respectively. Filled circles refer to micromagnetic simulation results.}
    \label{fig1}
\end{figure*}

The first considered magnetic system is the magnonic waveguide shown in fig.\ref{fig1}(a). The analysis of such a system has been inspired from ref.\cite{Wang_NatElectron2020}, where a YIG magnonic waveguide has been studied and proposed as directional coupler to build a magnonic logic circuit. 
The material parameters used are: saturation magnetization $M_s = 133\,$kA/m, damping $\alpha_\mathrm{G} = 2\times10^{-4}$ and exchange stiffness $A_\mathrm{ex} = 3.5\,$pJ/m. The micromagnetic equilibrium $\bm m_0$ is obtained by letting magnetization relax from a saturated state under a dc magnetic field of 50 mT applied along the same $x$ direction. 

The computation of eigenmodes $\bm\varphi_h$ and related eigenfrequencies $\omega_h$ provides in itself several insightful information, which are reported in fig.\ref{fig1}(a). One can see that the system exhibits modes with spatial distribution confined in a small region of the magnonic waveguide (e.g. mode labelled as 7) and modes which extend throughout the guide in different ways (e.g. modes 64, 73, 95). In particular, it is very interesting to notice that selected modes with distinct eigenfrequencies extend differently across the upper and lower arms in the rightmost part of the waveguide. Specifically, there are modes, such as mode 73 (mode 95) extending only in the right upper (lower) arm of the structure, and modes such as the 64 oscillating in both arms. These macro features are shared by several other modes associated with different eigenfrequencies. From the knowledge of the eigenpairs $(\omega_h,\bm\varphi_h)$, the dispersion curve of spin-waves propagating in the magnonic waveguide can be straightforwardly computed. The result is reported in in fig.\ref{fig1}(b). 
In fact, for each $\omega_h$ the two-dimensional Fourier Transform of the associated mode spatial distribution $\bm\varphi_h(\bm r)$ is computed in order to have its representation in the wavenumber domain $(k_x,k_y)$. Then, the spectral magnitude peaks along $k_y$ of the Fourier-transformed vector field  are plotted as function of $k_x$. By inspecting fig.\ref{fig1}(b), one can see that dominant spin-wave modes follow an approximately parabolic dispersion curve starting close to frequency values $\sim3.45\,$GHz. On one hand, this curve can qualitatively describe the oscillation pattern of spin-waves interpreted in terms of normal modes distribution, since confined (extended) spin-wave modes exhibit broad (narrow) spectra owing to Fourier Transform's duality property. On the other hand, the diagram is instrumental for analyzing the working condition of the magnonic waveguide. For instance, one can see that propagating extended modes with positive group velocity (narrow spectrum) are possible for positive $k_x$ starting  from $\omega/2\pi\sim3.6\,$GHz upwards. Vice versa, negative group velocities occur in the same range of frequencies, but for negative $k_x$ values.  
By using this approach, the dispersion relation is readily obtained once eigenmodes and associated eigenfrequencies, such as those reported in fig.\ref{fig1}(c), are computed. Moreover, this analysis immediately permits to reveal the presence of non-reciprocal propagation of spin waves \cite{Jamali_2013}. This is a great advantage with respect to its evaluation based on long micromagnetic simulations of magnetization dynamics driven by external excitations with suitable space-time profile. 
In this respect, we remark that the computation of the dispersion relation based on normal modes does not involve the knowledge of the external excitation, which is not the case of time-domain techniques.
In ref.\cite{Wang_NatElectron2020}, magnetization dynamics is excited in the magnonic waveguide by applying a microwave external field with a U-shaped antenna located in the initial part of the magnonic waveguide. Here we reproduce a similar scenario by applying a sinusoidal time-varying magnetic field with (normalized) amplitude $h_\mathrm{rf}$ and frequency $\omega_\mathrm{rf}$, directed along the $y$ direction, spatially-uniform in a region $1\,\mu \mathrm{m}$-long of the upper left arm of the waveguide. For weak  amplitudes of the microwave field (in dimensionless units $h_\mathrm{rf}\ll\alpha_\mathrm{G}$), magnetization dynamics can be analyzed in the linear regime, where the NMM can be written as: 
\begin{equation}\label{eq:lin_norm_modes_model}
\dot{a}_h = \frac{j\omega_h}{1+\alpha_\mathrm{G}^2}\, \left( b_{h} + \sum_i b_{hi}a_i\right)\,,
\end{equation}
where $b_h \approx -(\bm\varphi_h,\bm f_c)= B_h\,\cos(\omega_\mathrm{rf}t)$ ($B_h$ is the projection of the external field spatial distribution on $\bm \varphi_h$). It turns out that terms $b_{hi} = \delta_{hi} + b_{hi}^\alpha + b_{hi}^f$ are such that $|b_{hi}^\alpha/\alpha_\mathrm{G}|\ll1$ for $h\neq i$, $|b_{hh}^\alpha/\alpha_\mathrm{G}|\sim 1$, and all the terms $|b_{hi}^f/\alpha_\mathrm{G}|\ll 1$. For excitation frequency $\omega_\mathrm{rf}$ less than twice the value of the frequency corresponding to the first normal mode, parametric resonances can be excluded. This permits to consider a diagonal approximation of equations \eqref{eq:lin_norm_modes_model} where $b_{hi} \approx \,\delta_{hi}(1 + b_{hi}^\alpha)$. In this way, we can give a closed form expression for the normal modes dynamics as in the following equation:
\begin{equation}\label{eq:lin_norm_modes_evol}
\begin{aligned}
&a_h = A_h\, \left( e^{-\frac{t}{\tau_h}}(\cos{\omega_ht} + j\sin{\omega_h t})  +\frac{\omega_\mathrm{rf}}{j\omega_h-1/\tau_h}\sin{\omega_\mathrm{rf}t} - \cos{\omega_\mathrm{rf}t}\right)\,,\\
&A_h = \frac{-\omega_h^2\,B_h}{\omega_\mathrm{rf}^2-\omega_h^2-2j\,\omega_h/\tau_h}\,,\,\,\tau_h = |b_{hh}^\alpha|^{-1}=(\alpha_\mathrm{G}\omega_h^2\|\bm \varphi_h\|^2)^{-1} \,.
\end{aligned}
\end{equation}
where the initial condition $a_h(0) = 0$ is assumed.
The predictions of this formula in terms of $|a_h(t)|$ are compared with the results of micromagnetic simulations\cite{dAquino_JCP2005} in fig.\ref{fig1}(d), where the microwave field amplitude is $\mu_0 H_\mathrm{rf} = 10\,\mu$T and its frequency $f$ is equal to mode 95 frequency $f_{95}=3.946\,$GHz, which only extends in the lower arm of the waveguide (see the bottom of fig.\ref{fig1}(a)). This situation is useful to elucidate general properties of the NMM. In fact, when $f=f_h$, the denominator of the term $|A_h|$ in equation \eqref{eq:lin_norm_modes_evol} is minimized and the amplitude 
$|a_{h}(t)| = (1/2)\,\omega_\mathrm{rf}\tau_{h}\,|B_{h}|(1-e^{-\frac{t}{\tau_{h}}})$
has the typical aspect of first-order linear dynamical systems forced response. The maximum amplitude that is reached in steady-state conditions is proportional to $\tau_h\sim 1/\alpha_\mathrm{G}$ and to $b_{h}$. The latter coefficient $b_{h}$ has a very important meaning in that it quantifies the strength of the coupling between the spatial distribution of the $h$-th mode and the excitation field, as it can be inferred from its expression (see table \ref{tab:NMM} and section Methods). From the knowledge of the desired mode profile to excite, by using the latter formula, one can accordingly tune the external microwave  field direction and localization in order to maximize the excitation of the selected mode. These considerations are independent from the particular mode considered here and, therefore, hold in general for modes with different spatial distributions.

Thus, in the linear regime (low microwave power), there are two selection mechanisms for magnetization oscillations: i) selection due to the excitation field spatial distribution; ii) selection due to the excitation field frequency.
These mechanisms can be used to interpret the results reported in ref.\cite{Wang_NatElectron2020} where, depending on the frequency, magnetization oscillations were observed to occur either in one arm (upper or lower) or in both.
Indeed, if one considers a situation where the coupling of the first kind (spatial distribution) is of the same order of magnitude for modes in a given frequency range, then the coupling of the second kind (frequency) can be essential to produce larger magnetization oscillations in one arm (e.g. the lower one) rather than in the other (e.g. upper) arm. Such a difference can be quantitatively appreciated by introducing the following expression for the waveguide output power:
\begin{equation}\label{eq:rel_pow_arm_lin}
P_\mathrm{up} = \frac{\iint_{T\times\Omega_\mathrm{up}}\|\delta\bm m_\perp\|^2\,dV\,d\tau}{\iint_{T\times\Omega_\mathrm{up}}\|\delta\bm m_\perp\|^2\,dV\,d\tau + \iint_{T\times\Omega_\mathrm{lw}}\|\delta\bm m_\perp\|^2\,dV\,d\tau}\,,\hspace{0.5 cm} P_\mathrm{lw} = 1 - P_\mathrm{up}\,,
\end{equation}
where $\Omega_\mathrm{up(low)}$ indicate the final 1$\mu$m-long part of the upper (lower) arm. In equation \eqref{eq:rel_pow_arm_lin} magnetization oscillation is expressed in terms of the normal modes as $\delta\bm m_\perp(\bm r,t) = \sum a_k(t)\bm \varphi_k(\bm r)$ (see the section Methods) and $T$ is the period of the excitation field. The time-average is computed starting at $t = 1000\,$ns when the transients are reasonably extinguished (see. fig.\ref{fig1}(d)). In figure \ref{fig1}(e), the dependence of excitation field frequency on the relative power in each arm is plotted according to equation \eqref{eq:rel_pow_arm_lin}. One can easily see that, in the considered frequency range, there are multiple transitions of output power from the lower to the upper arm (and vice-versa) with different efficiency. 
In the linear regime, as it is expected from the above discussion according to equation \eqref{eq:rel_pow_arm_lin} and to the linear dependence of steady-state normal modes amplitudes on the microwave field magnitude (see eq.\eqref{eq:lin_norm_modes_evol}), the field frequency is the only parameter able to switch the output power between upper and lower arms. By increasing the microwave power, it is experimentally shown in ref.\cite{Wang_NatElectron2020} that the switching of output power between arms can be achieved as function of the microwave field amplitude, as a consequence of nonlinearity. 
In order to analyze this effect with the NMM, we consider eq.\eqref{eq:norm_modes_model} including terms up to the fourth order in mode amplitudes, while considering five normal modes (modes 58-62) whose frequencies ($f_{58},f_{59},f_{60},f_{61},f_{62}=3.6168,3.6186,3.6226,3.6267,3.6322$ GHz) are centered around the chosen microwave field frequency $f = 3.618\,$GHz.
%
In the linear regime, for such frequency, the relative power exhibits a sharp transition of output power between the arms (high sensitivity in frequency) and, therefore, it is expected that by inducing a nonlinear frequency shift, one can observe similar power switching between the waveguide arms. The NMM eq.\eqref{eq:norm_modes_model} with the above five modes is integrated for 1000 ns with different amplitudes of the excitation field. As a result of these computations, the relative power is evaluated as a function of the microwave amplitude according to eq.\eqref{eq:rel_pow_arm_lin} and plotted using solid lines in fig.\ref{fig1}(f). 
The diagram shows a dependence of the relative power on the microwave field amplitude which is very similar to the one reported in ref.\cite{Wang_NatElectron2020}. In the same figure, the results of full micromagnetic simulations are shown (filled dots). The agreement is quantitative and improves as far as the field amplitude increases. The NMM predicts a power switching field value (upper arm output power equals lower arm's one) $\sim 170\,\mu$T, while from micromagnetic simulations we have $\sim 70\,\mu$T. It is expected that this discrepancy can be reduced by increasing the number of modes considered in the integration of the NMM. In figure \ref{fig1}(f), the power vs microwave amplitude curves obtained from the NMM integrated with  20 modes (modes 41-60) are also plotted (dashed lines), which confirm the expected improvement of the agreement with micromagnetic simulations. 
We observe that, despite the limited number of modes considered, this analysis is in good agreement with micromagnetic simulations when the magnetization dynamics is in nonlinear regime and driven by moderate microwave field amplitudes $\mu_0H_\mathrm{rf}\sim 0.1\,$mT. As soon as the field amplitudes is increased to $\mu_0H_\mathrm{rf}\sim1\,$mT, the number of modes involved in the nonlinear dynamics grows dramatically with the microwave field power, leading to a chaotic regime. 

A systematic analysis of which modes are dominant in weakly nonlinear conditions is a topic of great interest for dimensionality reduction of nonlinear magnetization dynamics in magnonic systems and will be the focus of future investigation based on the NMM.

\subsection{High-power ferromagnetic resonance in rectangular magnetic nanodot}
In this section, the magnetic system considered is a rectangular thin-film with dimensions $L_\mathrm{x}=100\,$nm, $L_\mathrm{y}=50\,$nm and $L_\mathrm{z} = 3\,$nm. The material parameters are $M_\mathrm{s} = 800000$ A/m, $\alpha_\mathrm{G} = 0.02$, $l_\mathrm{ex}=5.69$ nm. The micromagnetic equilibrium is a quasi-uniform S-state along the $x$ direction obtained by relaxing a spatially-homogeneous magnetization slightly tilted in the transverse direction $y$. 
The rectangular nanodot magnetized at remanence is excited by a spatially-uniform and sinusoidally time-varying magnetic field applied along the $y$ direction with given amplitude and frequency to probe the steady-state response of magnetization oscillations. This situation is investigated with the NMM and compared with micromagnetic simulations\cite{dAquino_JCP2005} both performed by changing amplitude and frequency of the driving field as it is done in experiments to characterize the nonlinear FMR frequency response of the magnetic system. 
The NMM considered for the analysis of the above situation is obtained retaining the summation terms involving the product of up to 4 amplitudes of normal modes (see table \ref{tab:NMM} and section Methods). The integration of the modes nonlinear dynamics described by eq.\eqref{eq:norm_modes_model} is carried out by considering the first five modes. The initial magnetization state is the aforementioned equilibrium S-state, which is perturbed by the microwave field with constant amplitude and  frequency varying in the range $2-5$ GHz. The numerical simulations are performed by linearly  increasing (with a prescribed rate) the microwave frequency from the lowest to the largest value and decreasing it back to the lowest one in order to discriminate hysteretic effects in the frequency response as function of microwave field amplitude. In this respect, to rule out dynamical artifacts in the response curve related to the increasing/decreasing frequency rate of change, preliminary simulations have been performed to establish that no appreciable variation of results occurs below a value $\pm 0.01$ GHz/ns, which is consequently used to obtain the results presented in the sequel. 
\begin{figure*}
    \centering
    \includegraphics[width = 17 cm]{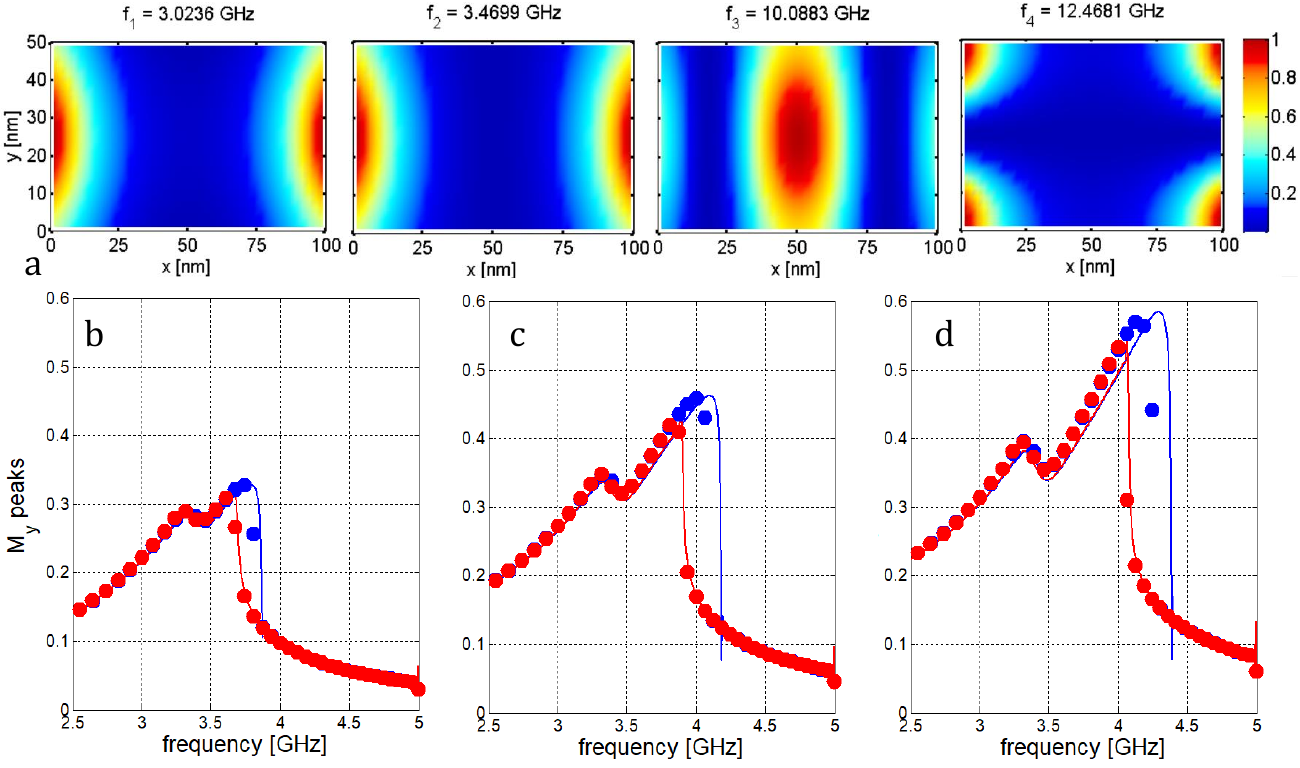}
    \caption{Normal mode analysis for the rectangular nanodot shown in fig.\ref{fig:geom_3sys}-(b).(a) Plots of the function $\|\bm \varphi\|^2$, normalized in the range $(0,1)$, for the first 4 normal modes. (b,c,d)  Steady-state oscillation amplitude of spatially-averaged magnetization $y$-component for three microwave field amplitudes: (b) $\mu_0 H_\mathrm{rf} = 1\,$mT, (c) $\mu_0 H_\mathrm{rf} = 1.5\,$mT, (d) $\mu_0 H_\mathrm{rf} = 2\,$mT. Solid lines refer to NMM, filled dots are the result of micromagnetic simulations. Blue (red) color refers to increasing (decreasing) frequency.}
    \label{fig2}
\end{figure*}

The comparison of the results of the NMM and micromagnetic simulations has been done in terms of average magnetization component. In particular, from the $y$-component of the spatially-averaged magnetization vector, the amplitude of steady-state oscillation has been considered.
In figure \ref{fig2}, such a comparison is shown for three amplitudes of the microwave field: (b) $\mu_0H_\mathrm{rf} = 1$ mT, (c) $\mu_0H_\mathrm{rf} = 1.5$ mT and (d) $\mu_0H_\mathrm{rf} = 2$ mT. First, one can clearly see that there is quantitative agreement between the NMM and full micromagnetic simulations. Second, it is apparent that, as far as the microwave amplitude is increased, more and more pronounced hysteresis in frequency response of magnetization oscillation appears, which is a strong evidence of nonlinear FMR. These results point out the capability of NMM of capturing strongly nonlinear magnetization dynamics with a very small number of degrees-of-freedom.
It is worthwhile remarking that, for uniformly magnetized nanoparticles, nonlinear FMR is amenable of analytical treatment in terms of bifurcation theory\cite{dAquino_TMag2017}.
Interestingly, despite the magnetic thin-film is almost uniformly magnetized, the frequency response exhibit multiple peaks ($\sim 3.2\,$GHz and $\sim 4\,$Ghz), which is the fingerprint of coexisting resonances. The resonance peaks are located at frequency values close to those of the first and second normal modes. From figure \ref{fig2}(a,b), one can see that such modes have a similar amplitude spatial distribution and, according to what has been discussed in the previous section for the linear regime, they can couple with the external microwave field by matching the spatial distribution and the frequency. 
When the system is driven in nonlinear conditions by increasing the microwave amplitude, the interpretation of the hysteretic behavior of the frequency response is rather complex to infer from micromagnetic simulations, whereas it becomes remarkably simple in terms of normal modes.
On the other hand, the uniform mode (macrospin) theory is not able to describe the observed dynamics, while the NMM with 5 degrees-of-freedom leads to quantitative estimation, providing also a way to extend the low-dimensional analysis of nonlinear magnetization oscillations to magnetic systems non uniformly magnetized.

By further increasing the field amplitude, the deviation of NMM from micromagnetic computations increases. This is due to the small number (five) of normal modes adopted to describe the magnetization dynamics. In particular, as far as the microwave amplitude is increased, a larger number of modes is required in eq.\eqref{eq:norm_modes_model} for the correct description of the dynamics. In this respect, the proposed NMM provides a class of low-dimensional dynamical models with increasing accuracy and predictive power comparable to full-scale micromagnetics.

\subsection{Synchronization in injection-locked Spin-Torque Oscillator}

\begin{figure*}
    \centering
    \includegraphics[width = 18 cm,height = 12 cm]{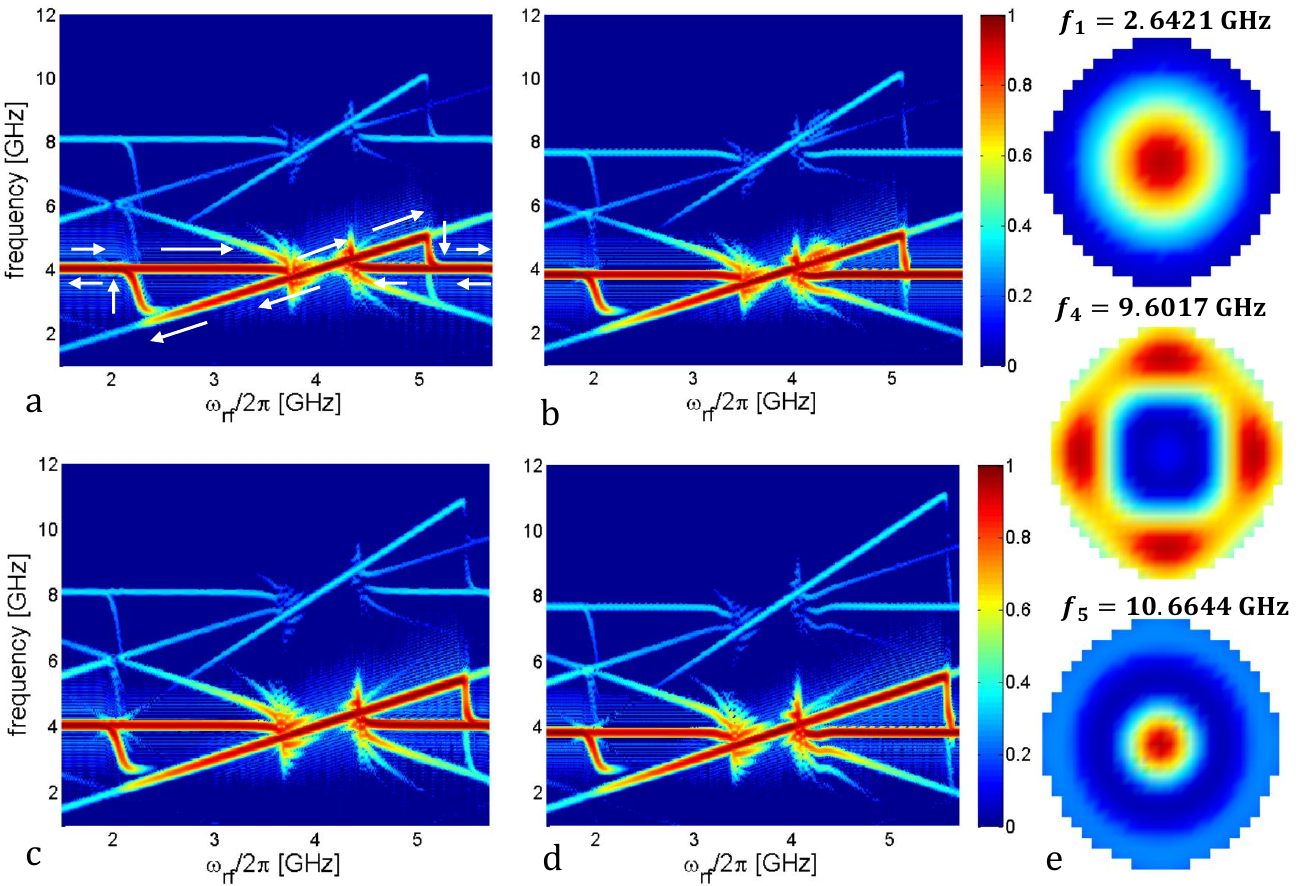}
    \caption{Normal mode analysis for the spintronic oscillator shown in fig.\ref{fig:geom_3sys}-(c). Comparison of the spectrograms of the in plane component of the average magnetization vector obtained for two distinct ac current values $J_{ac} = 7.33$ MA/cm$^2$ and $J_{ac} = 11.73$ MA/cm$^2$: (a,c) NMM and (b,d) micromagnetic simulations. (e) Plots of the function $\|\bm \varphi\|^2$, normalized in the range $(0,1)$, for the 3 normal modes (out of the 5 considered in the integration of the NMM) mainly involved in the synchronization dynamics. }
    \label{fig3}
\end{figure*}

In this section, the phase-locking of magnetization self-oscillations with an external microwave current is investigated for a spin transfer-torque nano-oscillator (STNO).
The magnetic system under investigation is a thin-disk with diameter $D = 100\,$nm and thickness $d = 12\,$nm. The material parameters are $M_s = 800000$ A/m, $\alpha_\mathrm{G} = 0.02$, $l_\mathrm{ex}=5.69$ nm. The micromagnetic equilibrium is obtained by relaxing magnetization under a spatially-uniform dc magnetic field $H_\mathrm{a} = 0.8\,M_s$ along the out-of-plane $z$-direction which is the symmetry axis of the disk. This produces a quasi-uniform magnetization aligned with the $z$ axis. In figure \ref{fig3}(e), three normal modes computed around this equilibrium state are reported.  
Magnetization dynamics is excited in the STNO by injecting a spin-polarized current in the perpendicular direction with dc+ac components, expressed in dimensionless form as: $\beta = \beta_{dc} + \beta_{ac} \cos(\omega_\mathrm{rf}t)$. The current value in dimensional units, can be obtained with the following expression: $J = \beta J_p /(2\eta)$ A/m$^2$, where $J_p = e\gamma M_s^2 L_z /(g\mu_B)$ with $\gamma$ the absolute value of the gyromagnetic ratio, $e$ is the absolute value of the electron charge, $g$ is the Landé factor, and $\mu_B$ is the Bohr magneton. The coefficient $\eta$ is the polarization factor, and is assumed to be $\eta = 0.1$ \cite{dAquino2020}.
The ac current amplitude is much smaller than the dc one. The spin-polarizer of the electric current is directed 45 degrees off the out-of-plane $z$ axis. 
In this situation, the dc spin-polarized current excites magnetization self-oscillations while the weak ac current is used to synchronize such oscillations with the external excitation, which is referred to as injection-locking. This is realized by slowly varying the ac excitation frequency back and forth in a range centered around the unperturbed self-oscillations (free running) frequency, In addition, for sufficiently large ac excitation, hysteretic synchronization in injection-locking is expected.
Here we analyze such a complex nonlinear phenomenon by using the proposed NMM with five normal modes.

In figure \ref{fig3}, the comparison of (normalized) power spectrograms of the $y$ component of the spatially-averaged magnetization vector, obtained from the NMM (see fig.\ref{fig3}(a,c)) and micromagnetic simulations\cite{dAquino_JCP2005} (see fig.\ref{fig3}(b,d)), are reported.
The dc spin polarized current density value is $J_{dc} = 29.34$ MA/cm$^2$, while the ac current is $J_{ac} = 7.33$ MA/cm$^2$ for fig.\ref{fig3}(a,b) and $J_{ac} = 11.73$ MA/cm$^2$ for fig.\ref{fig3}(c,d). In dimensionless units, they correspond to $\beta_{dc} = 0.2\,\alpha_\mathrm{G}$ and $\beta_{ac}=0.05\,\alpha_\mathrm{G}\,,\,0.08\,\alpha_\mathrm{G}$, respectively.
The free-running magnetization self-oscillation frequency is $\sim 4\,$GHz and the excitation frequency is linearly varied upwards and downwards in the range $1-6\,$ GHz with a rate of change $\pm 0.01\,$GHz/ns. No significant variation of the results for lower rate is observed.
In order to outline the synchronization mechanism, we consider the spectrogram reported in fig.\ref{fig3}(a). The simulated experiment starts from an unsynchronized state (US) oscillating at the unperturbed self-oscillation frequency (USF). Then, the excitation frequency is increased from 1 to 6 GHz. White arrows in the figure panels help understanding the synchronization process. In fact, magnetization oscillation frequency coincides with the free-running frequency before approaching a value of approximately $4\,$GHz. In this condition, the oscillation gets phase-locked to a synchronized state (SS) which is kept until the excitation frequency value reaches around 5 GHz, after which magnetization oscillations frequency drops at the USF. This abrupt change of frequency in terms of oscillating states correspond to the transition SS$\rightarrow$US. 
At this point, similar considerations hold for the reverse experiment where the excitation frequency is decreased back to 1 GHz (with with same rate $0.01\,$GHz/ns), but the synchronization band is different with respect to the forward experiment and the transition SS$\rightarrow$US corresponds to an upward jump.
From these diagrams, the hysteretic nature of the synchronization process can be immediately recognized. Indeed, for excitation frequency ranges (phase-locking bands) there is coexistence of SS and US. The actual observed magnetization regime depends on the past history of the dynamics.

Hysteretic synchronization of magnetization oscillations in a STNO has been theoretically investigated in previous works where the magnetization field distribution in the free layer of the STNO was spatially-uniform \cite{Serpico_TMag2009} and in a vortex state \cite{Perna_SciRep2016, dAquino_IEEEML2017}. In both cases, the magnetization dynamics was described by models with 2 degrees-of-freedom (collective variables).
For the STNO considered in this paper, the above approaches cannot be used to obtain quantitative predictions because the magnetization state is neither spatially-uniform nor in a vortex state. Nevertheless, even in this intermediate situation, the NMM is able to provide a low-dimensional dynamical model with predictive power comparable with full-scale micromagnetic simulations. In this respect, fig.\ref{fig3}(e) reports three of the five normal modes considered in the NMM which play the main role in the injection-locking of the STNO. 

\section{Discussion}
The generality of the proposed approach based on the description of nonlinear magnetization dynamics in terms of the normal modes associated with a micromagnetic equilibrium  has been established by a quantitative reproduction of the results of micromagnetic simulations for three different systems of great interest for magnetic technologies, subject to different external excitations (magnetic fields and spin-polarized currents). 
The main advantage of using such a model is twofold. One one hand, it allows to study the nonlinear magnetization dynamics with a strongly reduced number of degrees-of-freedom with respect to full micromagnetic simulations. For instance, the analysis of the magnonic waveguide in section \ref{sec:magnonic waveguide} required discretization on $N_c=N_x\times N_y\times N_z = 650\times62\times 1= 40300 $ computational cells ($3N_c \sim 10^5$ degrees-of-freedom), while the NMM allowed quantitative and increasingly accurate description of the relevant features of the nonlinear dynamics with $N_m=5$ and $N_m=20$ normal modes.  
Moreover, since normal modes are independent from the external  (forcing) excitation, the NMM can be used to perform fast engineering and optimization of microwave external fields and/or spin current on the basis of the desired feature (e.g. output power switch from upper to lower arms of the waveguide), which conversely would require repeating very long micromagnetic simulations (thousands nanoseconds to reach the steady-state) as far as the excitation is changed.
On the other hand, the NMM is a nonlinear system of ordinary differential equations which is much more manageable with analytical techniques. Indeed, in the case of linear magnetization dynamics, we have derived the expressions governing the time evolution of modes and, consequently, that of the magnetization field. Such formulas are independent from the magnetic system considered. Indeed, they have been derived independently from the magnetic equilibrium or the normal modes spatial profile. 
When nonlinear mode coupling  is considered in the dynamics, the NMM includes a large number of terms involving products of 2,3,... normal modes amplitudes. These terms become more important as the magnetic state deviates by a large angle from the equilibrium configuration. In this respect, it is worth to point out that, when the deviation of magnetization from the equilibrium becomes very large, as it may be the case of highly nonlinear processes such as magnetization switching, although the NMM could be in principle used, it would become rather soon impractically expensive in terms of computational cost. In fact, it would require to include a huge (ideally infinite) number of terms in both magnetization expansions \eqref{eq:magn_exp} and \eqref{eq:deltam_modes}, which means on one hand going far beyond the parabolic approximation of the unit-sphere, and on the other hand using a large number of eigenmodes. The a-priori estimation of the number of eigenmodes required to guarantee a prescribed accuracy in the description of magnetization dynamics is a very interesting topic and will be the focus of future investigations.
Nevertheless, there are many situations relevant for the design of magnetic technologies, such as those analyzed in this paper, where the NMM can quantitatively describe nonlinear magnetization dynamics by using a small number of modes, becoming a powerful alternative tool to full-scale micromagnetic simulations. In addition, the NMM allows to define a hierarchy of models which differ by the number of normal modes considered and type of nonlinear mode coupling (product of multiple modes amplitudes).
In conclusion, we believe that the proposed model will change the current perspective of computational micromgnetics and will help the design and optimization of current and novel  magnetic nanotechnologies.

\section{Methods}

\subsection{Normal Modes Formulation - conservative dynamics in the linear regime}\label{sectionA}

In this section, a brief review of the formulation of normal modes in micromagnetic systems is given. For details, the reader can refer to ref.\cite{dAquino_JCP2009}.\\
Let us consider a magnetic body occupying the region $\Omega$ whose magnetization field is denoted as $\bm M(\bm r,t)$. For a given temperature, well below the Curie temperature of the material, the fundamental micromagnetic constraint $|\bm M(\bm r,t)| = M_\mathrm{s}(T)$ is assumed, where $M_\mathrm{s}(T)$ is the saturation magnetization of the material at temperature $T$. By introducing the magnetization unit vector $\bm m(r,t)$, the micromagnetic constraint can be rewritten as:
\begin{equation}\label{eq:micromagnetic_constraint}
\bm m^2(\bm r,t)=1\,.
\end{equation}
In terms of energy, the magnetic state is characterized by the Gibbs-Landau free energy functional, which in dimensionless units can be written as:
\begin{equation}\label{eq:Gibbs_Landau_free_energy}
    g_L(\bm m; \bm h_\mathrm{a}) = \frac{G_L(\bm M;\bm H_\mathrm{a})}{\mu_0 M_\mathrm{s}^2 V_\Omega} = \frac{1}{V_\Omega}\int_\Omega\left[\frac{l_\mathrm{ex}^2}{2}(\nabla\bm m)^2-\frac{1}{2}\bm h_\mathrm{m}[\bm m]\cdot\bm m + \bm \kappa_\mathrm{an}[1-(\bm m\cdot \bm e_\mathrm{an})^2]-\bm h_\mathrm{a}\cdot\bm m\right]\,dV\,,
\end{equation}
where $\mu_0$ is the vacuum magnetic permeability, $V_\Omega$ is the volume of the magnetic body, $l_\mathrm{ex} = \sqrt{2A/(\mu_0M_s^2)}$ is the exchange length, $A$ is the exchange stiffness of the material, $\kappa_\mathrm{an},\,\bm e_\mathrm{an}$ are the anisotropy constant and direction respectively, $\bm h_a$ is the applied magnetic field and $\bm h_\mathrm{m}[\bm m]$ is the magnetostatic field, expressed in operator form as follows:
\begin{equation}
    \bm h_\mathrm{m}[\bm m]=-\nabla\nabla\cdot\frac{1}{4\pi}\int_\Omega \frac{\bm m(\bm r')}{|\bm r-\bm r'|}\,dV_{r'}\,.
\end{equation}
The minimization of Gibbs-Landau free energy functional \eqref{eq:Gibbs_Landau_free_energy} constrained on the unit-sphere \eqref{eq:micromagnetic_constraint} allows to derive the equations for micromagnetic equilibria, referred to as Brown's equations \cite{Brown_mumag}:
\begin{equation}
    \bm m_0\times\bm h_\mathrm{eff}[\bm m_0] = \bm 0\,\,in\,\Omega,\,\,\frac{\partial \bm m_0}{\partial \bm n} = \bm 0\,\, on\,\partial \Omega,
\end{equation}
where $\bm n$ is the outward normal to the boundary of the magnetic body $\partial \Omega$, and 
\begin{equation}
    \bm h_\mathrm{eff}[\bm m] = -\frac{\delta g_L}{\delta\bm m} = -\mathcal{C}\bm m+\bm h_\mathrm{a} = l_\mathrm{ex}^2\nabla^2\bm m + \bm h_\mathrm{m}[\bm m] +2 \kappa_\mathrm{an}(\bm m\cdot \bm e_\mathrm{an})\bm e_\mathrm{an} + \bm h_\mathrm{a}\,, 
\end{equation}
is the effective field given by the functional derivative with respect to the magnetization. The effective field is recast as the sum of a linear operator $-\mathcal{C}$ acting on the magnetization field plus the external applied field.\\
When the equilibrium is perturbed, magnetization dynamics occurs. For the time being, let us consider magnetization dynamics described by the conservative form of Landau-Lifshitz equation \cite{Landau1935}, which in dimensionless units reads as:
\begin{equation}\label{eq:conservative_LL}
    \frac{\partial \bm m}{\partial t} = -\bm m\times\bm h_\mathrm{eff}[\bm m]\,,
\end{equation}
where time is measured in units of $ (\gamma\,M_s)^{-1}$, with $\gamma$ being the absolute value of the gyromagnetic factor.
The magnetization field can be written in perturbation form as:
\begin{equation}\label{eq:pert_mag}
\bm m(\bm r,t) = \bm m_0(\bm r) + \delta\bm m(\bm r,t)\,\,.
\end{equation}
The micromagnetic constraint $|\bm m(\bm r,t)|^2=|\bm m_0(\bm r)+ \delta\bm m(\bm r,t)|^2=1$ implies that the following expansion (eq.\eqref{eq:magn_exp} in section Results, see next section for details) holds for magnetization:
\begin{equation}\label{eq:magn_exp2}
\bm m(\bm r,t) = \delta \bm m_\perp(\bm r,t) + \bm m_0(\bm r)\left(1-\frac{1}{2}\delta\bm m_\perp^2-\frac{3}{4!}\delta\bm m_\perp^4+\dots\right)\,,
\end{equation}
where $\delta \bm m_\perp(\bm r,t)$ is the component of $\delta\bm m$ pointwise perpendicular to the equilibrium $\bm m_0(\bm r)$. It is apparent that when $|\delta\bm m|\ll 1$, one can retain terms up to first order in eq.\eqref{eq:magn_exp2}, leading to a  perturbation field $\delta\bm m\approx \delta\bm m_\perp$ that lies in the plane perpendicular to the equilibrium. We denote the function space of square-integrable vector fields of this type as the tangent space $\mathcal{TM}(\bm m_0)$. 
The time-evolution of such small perturbation field can be then described by the linearized form of equation \eqref{eq:conservative_LL} around the equilibrium $\bm m_0$, which can be expressed as follows\cite{dAquino_JCP2009}:
\begin{equation}\label{eq:lin_cons_LL}
    \frac{\partial \delta\bm m_\perp}{\partial t} = \bm m_0\times(\mathcal{C}+h_0)\delta\bm m_\perp = \bm m_0\times \mathcal{A}_{0\perp}\delta\bm m_\perp \,,
\end{equation}
where $\mathcal{A}_{0\perp}=\mathcal{P}_{\bm m_0}(\mathcal{C}+h_0\,\mathcal{I})$, with $\mathcal{P}_{\bm m_0} = \mathcal{I} - \bm m_0\otimes\bm m_0$ the projection operator on plane orthogonal to $\bm m_0$, $\mathcal{I}$ the identity operator and $h_0 = \bm m_0 \cdot \bm h_\mathrm{eff}[\bm m_0]$.
If one searches for time-harmonic solutions of the type $\delta\bm m_\perp(\bm r,t) =$ Re$\left[\bm \varphi(\bm r)\,e^{j\omega t}\right]$ oscillating in each point around the ground state, which means using the phasor notation, eq.\eqref{eq:lin_cons_LL} can be recast as the following generalized eigenvalue problem\cite{dAquino_JCP2009}:
\begin{equation}\label{eq:gen_eig_prob}
    \mathcal{A}_{0\perp}\bm \varphi = \omega\, \mathcal{B}_0\bm\varphi \,,
\end{equation}
where $\mathcal{B}_0\bm\varphi = -j\bm m_0\times\bm\varphi$. The summary of the main properties of the above problem, which will be extensively used in the next sections, are the following (detailed derivations can be found in ref.\cite{dAquino_JCP2009}):
\begin{enumerate}
    \item\label{prop1} The eigenvalues $\omega_k$ (eigenfrequencies) are real.
    \item\label{prop2} If the eigenpair $(\omega_k,\bm \varphi_k)$ is solution of the problem \eqref{eq:gen_eig_prob}, then so is also the eigenpair $(-\omega_k,\bm \varphi_k^*)$, where $\varphi_k^*$ is the complex conjugate of $\varphi_k$.
    \item\label{prop3} The spectrum $\{\omega_k\}_{k\epsilon\mathbb{Z}}$ is discrete (the essential spectrum is at infinity), with $\omega_{-k}=-\omega_k$.
    \item\label{prop4} The set of eigenvectors (eigenmodes) $\{\bm\varphi_k\}_{k\epsilon\mathbb{Z}}$ constitute a basis of the tangent space $\mathcal{TM}(\bm m_0)$, with $\bm\varphi_{-k} = \bm\varphi_k^*$.
     \item\label{prop5}The eigenmodes of different and non degenerate eigenfrequencies are $\mathcal{A}_{0\perp}$-orthogonal. 
\end{enumerate}
According to property \ref{prop4} one can represent any vector field in $\mathcal{TM}(\bm m_0)$ and in particular $\delta\bm m_\perp$ as:
\begin{equation}\label{eq:prop4}
    \delta\bm m_\perp(\bm r,t) = \sum_{k=-\infty}^\infty a_k(t)\,\bm\varphi_k(\bm r)\,,
\end{equation}
while property \ref{prop5} can be explicitly specified as follows:
\begin{equation}\label{eq:A0_orth_1}
  (\bm\varphi_h,\bm\varphi_k)_{\mathcal{A}_{0\perp}}=(\bm\varphi_h,\mathcal{A}_{0\perp}\bm\varphi_k) =  \frac{1}{V_\Omega}\int_\Omega \bm\varphi_h^*\cdot\mathcal{A}_{0\perp}\bm\varphi_k\,dV =  \delta_{hk}\,, \end{equation}
  with $V_\Omega$ being the volume of the region $\Omega$ occupied by the magnetic body.
It is useful to remark that due to the self adjointness of the operator $\mathcal{A}_{0\perp}$, the equation above can be rewritten as:
\begin{equation}\label{eq:A0_orth_2}
  (\bm\varphi_h,\bm\varphi_k)_{\mathcal{A}_{0\perp}}=(\mathcal{A}_{0\perp}\bm\varphi_h,\bm\varphi_k) = j\omega_h(\bm m_0\times\bm\varphi_h,\bm\varphi_k) =  \delta_{hk}\,. 
\end{equation}
It is worthwhile to remark that the magnetization perturbation field $\delta\bm m(\bm r,t)$ associated to the $h$-th magnetic normal mode with eigenfrequency $\omega_h$ is real and can be expressed as $\delta\bm m_\perp(\bm r,t) = a_h(t)\,\bm\varphi_h(\bm r) + a_h^*(t)\,\bm\varphi_h^*(\bm r)=2\,$Re$\,[a_h(t)\,\bm\varphi_h(\bm r)]$.\\
Now, we substitute eq. \eqref{eq:prop4} in eq. \eqref{eq:lin_cons_LL}, and use the orthogonality of the eigenmodes \eqref{eq:A0_orth_1} as follows:
\begin{equation}\label{eq:proj_lin_cons}
    \begin{aligned}
    &\left(\bm \varphi_h,\frac{\partial \delta\bm m_\perp}{\partial t}\right)_{\mathcal{A}_{0\perp}} = \sum_k\left(\bm \varphi_h,\bm\varphi_k\right)_{\mathcal{A}_{0\perp}}\dot{a}_k = \dot{a}_h\,,\\
    & \left(\bm \varphi_h,\bm m_0\times \mathcal{A}_{0\perp}\delta\bm m_\perp\right)_{\mathcal{A}_{0\perp}} = \sum_k \left(\bm \varphi_h,\bm m_0\times \mathcal{A}_{0\perp}\bm \varphi_k\right)_{\mathcal{A}_{0\perp}} a_k = \sum_kj\omega_k\left(\bm \varphi_h,\bm\varphi_k\right)_{\mathcal{A}_{0\perp}}a_k = j\omega_h a_h\,.
    \end{aligned}
\end{equation}
Then, one can conclude that the micromagnetic dynamics is described by the following system of ordinary differential equations whose unknowns are the normal modes amplitudes:
\begin{equation}\label{eq:lin_norm_mod}
    \dot{a}_h = j\omega_h\,a_h\,.
\end{equation}
According to the latter equation, each amplitude  evolves independently from the others according to the following law: $a_h(t) = a_{h0}\,e^{j\omega_h (t-t_0)}$, with $a_{h0}=a_h(t = t_0)$ being the initial amplitude at time $t_0$. \\
As it will be shown in the next sections, when also material dissipation and external forcing are taken into account, even in the linear regime, coupling between different normal modes occurs. For moderately larger amplitude values, more pronounced coupling of nonlinear nature sets in.\\
In the sequel, we derive the general system of coupled differential equations for the complex amplitudes describing nonlinear magnetization processes. By neglecting nonlinear terms in such system, one can describe linear magnetization dynamics driven by generic external excitations in terms of mode amplitudes' evolution. For instance, this kind of analysis can provide the basis for investigations of magnonic devices (see section Results).

\subsection{Conservative Dynamics - non linear regime}

When the perturbation of the equilibrium state brings the magnetization dynamics out of the linear regime, the assumption of Section \ref{sectionA}  $\delta\bm m\approx \delta \bm m_\perp$ does not hold any longer and hence eq. \eqref{eq:lin_norm_mod} describing the normal modes amplitude evolution is no more valid. In the following, we arrive at a new system of ODEs which generalizes equation \eqref{eq:lin_norm_mod} for nonlinear magnetization dynamics. 
As a first step, we decompose the magnetization field according to the following formula:
\begin{equation}\label{eq:decomp_var_magn}
    \bm m(\bm r,t) = \bm m_0(\bm r) + \delta\bm m(\bm r,t) = \bm m_0(\bm r) + \delta\bm m_0(\bm r,t) + \delta\bm m_\perp(\bm r,t)\,,
\end{equation}
where for each point $\bm r\,\epsilon\,\Omega$, $\delta \bm m_0(\bm r,t) = \delta m_0(\bm r,t)\, \bm m_0(\bm r)$ is the projection of $\delta\bm m(\bm r,t)$ along the equilibrium $\bm m_0(\bm r)$.
Then, plugging equation \eqref{eq:decomp_var_magn} into the conservative LLG equation 
\eqref{eq:conservative_LL}, we obtain:
\begin{equation}\label{eq:nonlin_md_cons_1}
    \frac{\partial\delta\bm m}{\partial t} = \bm m_0\times(\mathcal{C}+h_0)\delta\bm m + \delta\bm m\times \mathcal{C}\,\delta\bm m\,,
\end{equation}
which can be expanded in the following equation:
\begin{equation}\label{eq:nonlin_md_cons_2}
\begin{aligned}
\bm m_0 \frac{\partial\delta m_0}{\partial t} + \frac{\partial\delta\bm m_\perp}{\partial t} =& \bm m_0\times\mathcal{A}_{0\perp}\delta\bm m_\perp + \bm m_0\times \mathcal{C}\,\delta\bm m_0 + \delta\bm m_0\times \mathcal{C}\,\delta\bm m_0 \\
&+\delta\bm m_\perp\times\mathcal{C}\,\delta\bm m_0 + \delta\bm m_0\times\mathcal{C}\,\delta\bm m_\perp+ \delta\bm m_\perp\times\mathcal{C}\,\delta\bm m_\perp\,,
\end{aligned}
\end{equation}
where we use the fact that $\bm m_0\times h_0\delta\bm m = \bm m_0\times h_0\delta\bm m_\perp$.
By decomposing the operator $\mathcal{C}$ as
\begin{equation}\label{eq:decomp_op_C}
  \mathcal{C} =  \mathcal{C}_\perp +\, \mathcal{C}_0\,,\hspace{1cm} \mathcal{C}_\perp = \mathcal{P}_{\bm m_0}\mathcal{C}\,,
\end{equation}
equation \eqref{eq:nonlin_md_cons_2} is also decomposed in the following system of coupled equations:
\begin{equation}\label{eq:nonlin_md_cons_3}
\begin{aligned}
\frac{\partial\delta\bm m_\perp}{\partial t} = & \bm m_0\times\mathcal{A}_{0\perp}\delta\bm m_\perp + \bm m_0\times \mathcal{C}_\perp\,\delta\bm m_0 + \delta\bm m_0\times \mathcal{C}_\perp\,\delta\bm m_0 + \\
&\delta\bm m_0\times\mathcal{C}_\perp\,\delta\bm m_\perp + \delta\bm m_\perp\times\mathcal{C}_0\,\delta\bm m_0 + \delta\bm m_\perp\times\mathcal{C}_0\,\delta\bm m_\perp\,,\\
\bm m_0\frac{\partial\delta m_0}{\partial t} = & \delta\bm m_\perp\times\mathcal{C}_\perp\,\delta\bm m_0 + \delta\bm m_\perp\times\mathcal{C}_\perp\,\delta\bm m_\perp\,.
\end{aligned}
\end{equation}
Up to this point, we have rewritten the conservative Landau-Lifshitz equation in the reference frame defined by the unit vector $\bm m_0$ and the plane perpendicular to $\bm m_0$. The system of ordinary differential equations in the $a_h(t)$ could be obtained by applying property \ref{prop5} to the first of equations \eqref{eq:nonlin_md_cons_3}. However, the occurrence of terms including $\delta m_0$ couples the projected equation with the second of equations \eqref{eq:nonlin_md_cons_3}. This can be circumvented by expressing $\delta m_0$ as a function of complex amplitudes $a_h(t)$ and eigenmodes $\bm\varphi_h(\bm r)$. Such a relationship exists as a consequence of the fundamental micromagnetic constraint imposed on equation \eqref{eq:decomp_var_magn}. 
Therefore, by substituting eq.\eqref{eq:decomp_var_magn} in eq.\eqref{eq:micromagnetic_constraint}, we obtain the following expressions:
\begin{equation}\label{eq:dm0_func_dmperp}
    \delta m_\perp^2+2\delta m_0+\delta m_0^2 = 0\,\Leftrightarrow\, \delta m_0 = -1 \pm \sqrt{1-\delta m_\perp^2}\,.
\end{equation}
If the magnetization dynamics occurs around the equilibrium $\bm m_0$, then the $+$ sign has to be selected. Indeed, when $\delta\bm m_\perp = \bm 0$ such sign choice produces $\delta m_0 = 0$. The right hand side of equation \eqref{eq:dm0_func_dmperp} can be expanded in Taylor series as follows:
\begin{equation}\label{eq:Taylor_series_dm0}
    \delta m_0 = -\frac{1}{2}\delta m_\perp^2-\frac{3}{4!}\delta m_\perp^4 + \dots\,,
\end{equation}
where by considering only the first term of the expansion and expressing $\delta \bm m_\perp$ as a function of the normal modes, we obtain the following equation:
\begin{equation}\label{eq:parab_approx_dm0}
    \delta m_0(\bm r,t) \approx -\frac{1}{2}\sum_{i,j = -\infty}^{+\infty}\bm\varphi_i(\bm r)\cdot\bm\varphi_j(\bm r)\,a_i(t)a_j(t) = -\frac{1}{2}\sum_{i,j = -\infty}^{+\infty}\psi_{ij}(\bm r)\,a_i(t)a_j(t)\,.
\end{equation}
This equation represents an approximation which will be referred to as parabolic approximation. It is quite accurate when $\delta m_0 < 0.2$, which corresponds to $\delta m_\perp < 0.6$. We will discover in the following that this range is enough to quantitatively describe nonlinear dynamics around the equilibrium state in most cases of interest for applications. Moreover, the accuracy and the range of validity of the model developed in the following can be increased by inserting further terms in equation \eqref{eq:parab_approx_dm0} coming from the Taylor series \eqref{eq:Taylor_series_dm0}.
The set constituted by equation \eqref{eq:parab_approx_dm0} and the property \ref{prop5} allows to derive the projection of the the first of equations \eqref{eq:nonlin_md_cons_3} in the normal mode basis as follows:
\begin{equation}\label{eq:proj_nonlin_cons}
\begin{aligned}
\left(\bm\varphi_h,\bm m_0\times \mathcal{C}_\perp\delta\bm m_0\right)_{\mathcal{A}_{0\perp}} = & j\omega_h\left(\bm\varphi_h, \mathcal{C}\delta\bm m_0\right) = -\frac{j\omega_h}{2}\sum_{i,j}\left(\bm\varphi_h, \mathcal{C}\psi_{ij}\bm m_0\right)a_ia_j\,,\\
\left(\bm\varphi_h,\delta\bm m_0\times \mathcal{C}_\perp\delta\bm m_0\right)_{\mathcal{A}_{0\perp}} = & \frac{j\omega_h}{4}\sum_{i,j,k,l}\left(\bm\varphi_h, \psi_{kl}\,\mathcal{C}\psi_{ij}\bm m_0\right)a_ia_ja_ka_l\,,\\
\left(\bm\varphi_h,\delta\bm m_0\times \mathcal{C}_\perp\delta\bm m_\perp\right)_{\mathcal{A}_{0\perp}} = & -\frac{j\omega_h}{2}\sum_{i,j,k,l}\left(\bm\varphi_h, \psi_{jk}\,\mathcal{C}\bm\varphi_i\right)a_ia_ja_k\,,\\
\left(\bm\varphi_h,\delta\bm m_\perp\times \mathcal{C}_0\delta\bm m_0\right)_{\mathcal{A}_{0\perp}} = & -j\omega_h \left(\bm\varphi_h,\bm m_0\cdot\mathcal{C}\delta\bm m_0\,\delta\bm m_\perp \right) =  \frac{j\omega_h}{2}\sum_{i,j,k}\left(\bm\varphi_h,\bm m_0\cdot\mathcal{C}\psi_{jk}\bm m_0\,\bm \varphi_i \right) a_ia_ja_k\,,\\
\left(\bm\varphi_h,\delta\bm m_\perp\times \mathcal{C}_0\delta\bm m_\perp\right)_{\mathcal{A}_{0\perp}} = &-j\omega_h \left(\bm\varphi_h,\bm m_0\cdot\mathcal{C}\delta\bm m_\perp\,\delta\bm m_\perp \right) =  -j\omega_h\sum_{i,j}\left(\bm\varphi_h,\bm m_0\cdot\mathcal{C}\bm\varphi_j\,\bm \varphi_i \right) a_ia_j\,.\\
\end{aligned}
\end{equation}
By putting equations \eqref{eq:proj_lin_cons} and \eqref{eq:proj_nonlin_cons}  together, we arrive at the following system of ODEs in the normal modes amplitudes:
\begin{equation}\label{eq:nonlin_cons_odes_mode}
\dot{a}_h = j\omega_h\,a_h + j\omega_h\sum_{i,j}c_{hij}^0 a_ia_j\,+ \frac{j\omega_h}{2}\sum_{i,j,k}d_{hijk}^0a_ia_ja_k\,+ \frac{j\omega_h}{4}\sum_{i,j,k,l}e_{hijkl}^0a_ia_ja_ka_l\,,
\end{equation}
where:
\begin{equation}\label{eq:coeff_odes_nonlin_cons_mode}
 \begin{aligned}
  &c_{hij}^0 = -\frac{1}{2}\left(\bm\varphi_h, \mathcal{C}\psi_{ij}\bm m_0\right)-\left(\bm\varphi_h,\bm m_0\cdot\mathcal{C}\bm\varphi_j\,\bm \varphi_i \right)\,,\\
 & d_{hijk}^0 = -\left(\bm\varphi_h, \psi_{jk}\,\mathcal{C}\bm\varphi_i\right) + \left(\bm\varphi_h,\bm m_0\cdot\mathcal{C}\psi_{jk}\bm m_0\,\bm \varphi_i \right)\,,\\
 & e_{hijkl}^0 = \left(\bm\varphi_h, \psi_{kl}\,\mathcal{C}\psi_{ij}\bm m_0\right)\,.\\
 \end{aligned}
\end{equation}
Then we arrived to an equivalent formulation of the conservative Landau-Lifshitz equation in terms of the normal modes amplitudes. The only approximation used, as discussed previously, is the parabolic approximation expressed by equation \eqref{eq:parab_approx_dm0}. 

\subsection{Damped Dynamics - non linear regime}
When the damping is included, magnetization dynamics is  assumed to be described by the Landau-Lifshitz-Gilbert equation \cite{BMS}, which in dimensionless units reads as:
\begin{equation}\label{eq:LLG}
    \frac{\partial \bm m}{\partial t} -\alpha_\mathrm{G}\bm m\times \frac{\partial\bm m}{\partial t}= -\bm m\times\bm h_\mathrm{eff}[\bm m]\,,
\end{equation}
where $\alpha_\mathrm{G}\sim 10^{-2}$ is the Gilbert damping. This equation can be put in the explicit form $\dot{\bm x} = \bm f[\bm x;t]$, obtaining the Landau-Lifshitz equation \cite{BMS}:
\begin{equation}\label{eq:damp_LL}
    (1+\alpha_\mathrm{G}^2)\,\frac{\partial \bm m}{\partial t} = -\bm m\times\bm h_\mathrm{eff}[\bm m]-\alpha_\mathrm{G}\,\bm m\times(\bm m\times\bm h_\mathrm{eff}[\bm m])\,.
\end{equation}
In order to include damping effects in the normal modes dynamics, we project the Landau-Lifshitz equation in the tangent space $\mathcal{TM}(\bm m_0)$ as follows:
\begin{equation}\label{eq:proj_LLeq_perp}
\begin{aligned}
(1+\alpha_\mathrm{G}^2)\frac{\partial\delta\bm m_\perp}{\partial t} = & \bm m_0\times\mathcal{A}_{0\perp}\delta\bm m_\perp + \bm m_0\times \mathcal{C}_\perp\,\delta\bm m_0 + \delta\bm m_0\times \mathcal{C}_\perp\,\delta\bm m_0 + \delta\bm m_0\times\mathcal{C}_\perp\,\delta\bm m_\perp + \\
&\delta\bm m_\perp\times\mathcal{C}_0\,\delta\bm m_0 + \delta\bm m_\perp\times\mathcal{C}_0\,\delta\bm m_\perp - \alpha_\mathrm{G}\,\bm m\times(\bm m\times\bm h_\mathrm{eff}[\bm m])\big{|}_\perp\,,
\end{aligned}
\end{equation}
where, according to equations \eqref{eq:decomp_var_magn} and \eqref{eq:decomp_op_C}, we have:
\begin{equation}
  \bm m\times(\bm m\times\bm h_\mathrm{eff}[\bm m])\big{|}_\perp = \bm m\cdot h_\mathrm{eff}[\bm m]\,\delta\bm m_\perp +\mathcal{C}_\perp(\delta\bm m_\perp + \delta\bm m_0)\,,
\end{equation}
with:
\begin{equation}
    \bm m\cdot \bm h_\mathrm{eff}[\bm m] = (1+\delta m_0)\left(h_0-\bm m_0\cdot\mathcal{C}(\delta \bm m_\perp+\delta \bm m_0)\right) - \delta\bm m_\perp\cdot\mathcal{C}(\delta\bm m_\perp + \delta\bm m_0)\,.
\end{equation}
By combining the two above equations, we arrive at the following equation:
\begin{equation}
\begin{aligned}
   \bm m\times(\bm m\times\bm h_\mathrm{eff}[\bm m])\big{|}_\perp =& \mathcal{A}_{0\perp} \delta\bm m_\perp + h_0\,\delta m_0\, \delta\bm m_\perp + \mathcal{C}_\perp\delta\bm m_0 - (1+\delta m_0)\,\bm m_0\cdot\mathcal{C}(\delta \bm m_\perp+\delta \bm m_0)\,\delta\bm m_\perp +\\
   &- \delta\bm m_\perp\cdot\mathcal{C}(\delta\bm m_\perp + \delta\bm m_0)\,\delta\bm m_\perp\,,
   \end{aligned}
\end{equation}
from which, similarly as we did for equations \eqref{eq:proj_nonlin_cons}, we can derive the projection of the damping torque in the normal modes basis expressed by the following terms:
\begin{equation}\label{eq:proj_nonlin_damp}
\begin{aligned}
&\left(\bm\varphi_h,\mathcal{A}_{0\perp} \delta\bm m_\perp\right)_{\mathcal{A}_{0\perp}} = \omega_h\sum_{i}\omega_i\,\left(\bm\varphi_h, \bm\varphi_i\right)\,a_i\,,\\
&\left(\bm\varphi_h,h_0\,\delta m_0\delta\bm m_\perp\right)_{\mathcal{A}_{0\perp}} =  -\frac{j\omega_h}{2}\sum_{i,j,k}\left(\bm m_0\times\bm\varphi_h,h_0\, \psi_{jk}\,\bm\varphi_i\right)\,a_ia_ja_k\,,\\
&\left(\bm\varphi_h,\mathcal{C}_\perp\delta\bm m_0\right)_{\mathcal{A}_{0\perp}} =  -\frac{j\omega_h}{2}\sum_{i,j}\left(\bm m_0\times\bm\varphi_h,\mathcal{C}\psi_{ij}\bm m_0\right)\,a_ia_j\,,\\
&\left(\bm\varphi_h,- (1+\delta m_0)\,\bm m_0\cdot\mathcal{C}\delta \bm m_\perp\,\delta \bm m_\perp\right)_{\mathcal{A}_{0\perp}} = -j\omega_h\sum_{i,j}\left(\bm m_0\times\bm\varphi_h,\bm m_0\cdot\mathcal{C}\bm\varphi_j\bm\varphi_i\right)\,a_ia_j\,+\\ &\hspace{5cm}\frac{j\omega_h}{2}\sum_{i,j,k,l}\left(\bm m_0\times\bm\varphi_h,\psi_{kl}\,\bm m_0\cdot\mathcal{C}\bm\varphi_j\bm\varphi_i\right)\,a_ia_ja_ka_l\,,\\
&\left(\bm\varphi_h,-\,\bm m_0\cdot\mathcal{C}\delta \bm m_0\,\delta \bm m_\perp\right)_{\mathcal{A}_{0\perp}} = \frac{j\omega_h}{2}\sum_{i,j,k}\left(\bm m_0\times\bm\varphi_h,\bm m_0\cdot\mathcal{C}\psi_{jk}\bm m_0\bm\varphi_i\right)\,a_ia_ja_k\,\\ 
&\left(\bm\varphi_h,- \delta m_0\,\bm m_0\cdot\mathcal{C}\delta \bm m_0\,\delta \bm m_\perp\right)_{\mathcal{A}_{0\perp}} =-\frac{j\omega_h}{4}\sum_{i,j,k,l,m}\left(\bm m_0\times\bm\varphi_h,\psi_{lm}\,\bm m_0\cdot\mathcal{C}\psi_{jk}\,\bm m_0\bm\varphi_i\right)\,a_ia_ja_ka_la_m\,,\\
&\left(\bm\varphi_h,- \delta\bm m_\perp\cdot\mathcal{C}\delta \bm m_\perp\,\delta \bm m_\perp\right)_{\mathcal{A}_{0\perp}} = -j\omega_h\sum_{i,j,k}\left(\bm m_0\times\bm\varphi_h,\,\bm \varphi_k\cdot\mathcal{C}\bm\varphi_j\,\bm\varphi_i\right)\,a_ia_ja_k\,,\\
&\left(\bm\varphi_h,- \delta\bm m_\perp\cdot\mathcal{C}\delta \bm m_0\,\delta \bm m_\perp\right)_{\mathcal{A}_{0\perp}} =\frac{j\omega_h}{2}\sum_{i,j,k,l}\left(\bm m_0\times\bm\varphi_h,\bm \varphi_l\cdot\mathcal{C}\psi_{jk}\bm m_0\,\bm\varphi_i\right)\,a_ia_ja_ka_l\,.\\
\end{aligned}
\end{equation}
At this point, we are ready to generalize the system of odes \eqref{eq:nonlin_cons_odes_mode} including damping terms. In this respect, we have:
\begin{equation}\label{eq:nonlin_damp_odes_mode}
\begin{aligned}
(1+\alpha_\mathrm{G}^2)\,\dot{a}_h = &j\omega_h\sum_i b_{hi}a_i + j\omega_h\sum_{i,j}c_{hij} a_ia_j\,+ \frac{j\omega_h}{2}\sum_{i,j,k}d_{hijk}a_ia_ja_k\,+\\
&\frac{j\omega_h}{4}\sum_{i,j,k,l}e_{hijkl}a_ia_ja_ka_l\,+ \frac{j\omega_h}{4}\sum_{i,j,k,l,m}f_{hijklm}a_ia_ja_ka_la_m\,,
\end{aligned}
\end{equation}
where:
\begin{equation}\label{eq:coeff_odes_nonlin_damp_mode}
 \begin{aligned}
 &b_{hi} = \delta_{hi}+\alpha_\mathrm{G}\,b_{hi}^\alpha\,,\\
 &c_{hij} = c_{hij}^0 +\alpha_\mathrm{G}\,c_{hij}^\alpha\,,\\
 & d_{hijk} = d_{hijk}^0+\alpha_\mathrm{G}\,d_{hijk}^\alpha \,,\\
 & e_{hijkl} = e_{hijkl}^0+\alpha_\mathrm{G}\,e_{hijkl}^\alpha\,.\\
 & f_{hijklm} = \alpha_\mathrm{G}\,f_{hijklm}^\alpha\,,\\
 \end{aligned}
\end{equation}
with $b_{hi}^\alpha,\,c_{hij}^\alpha,\,d_{hijk}^\alpha,\,e_{hijkl}^\alpha,\,f_{hijklm}^\alpha$ being the coefficients inferred from equation \eqref{eq:proj_nonlin_damp} which multiply the product of 1, 2, 3, 4 and 5 normal modes amplitudes, respectively.

\subsection{Forced Dynamics - non linear regime}

The last step for the study of magnetization dynamics in the framework of normal modes is the inclusion of external excitation terms in the model. Two different kinds of excitation will be considered: magnetic field and spin-polarized current. In general, no restriction on spatial distribution and time dependence is required. The spin-transfer torque term will be assumed of the Slonczewski type. In this respect, the equation describing LLGS magnetization dynamics is the following\cite{BMS}:
\begin{equation}\label{eq:LLGS}
    \frac{\partial \bm m}{\partial t} -\alpha_\mathrm{G}\bm m\times \frac{\partial\bm m}{\partial t}= -\bm m\times\bm h_\mathrm{eff}[\bm m]-\bm m\times\bm h_\mathrm{rf} +\beta\bm m\times(\bm m\times\bm p)\,,
\end{equation}
where $\bm h_\mathrm{rf}$ is the normalized microwave magnetic field and $(\beta,\bm p)$ is the pair identifying the normalized amplitude and the polarization unit vector of the spin current. As we did in the previous section, it is useful to rewrite equation \eqref{eq:LLGS} in explicit form as follows:
\begin{equation}\label{eq:LLGS_canform}
    (1+\alpha_\mathrm{G}^2)\,\frac{\partial \bm m}{\partial t}= -\bm m\times\bm h_\mathrm{eff}[\bm m] -\alpha_\mathrm{G}\,\bm m\times(\bm m\times\bm h_\mathrm{eff}[\bm m])-\bm m\times\bm f_c -\bm m\times(\bm m\times\bm f_d)\,,
\end{equation}
where $\bm f_c = \bm h_\mathrm{rf} +\alpha_\mathrm{G}\beta\bm p$ and $\bm f_d =  \alpha_\mathrm{G}\,\bm h_\mathrm{rf} -\beta\,\bm p $.
At this point, we can derive the several terms to be included in the system of ODEs \eqref{eq:nonlin_damp_odes_mode}. Specifically, the terms due to $\bm f_c$ are expressed in the following:
\begin{equation}\label{eq:proj_nonlin_forc_1}
\begin{aligned}
    &\bm m\times\bm f_c = (1+\delta m_0)\,\bm m_0\times\bm f_c + \delta\bm m_\perp\times\bm f_c\,,\\
    &\left(\bm\varphi_h, (1+\delta m_0)\,\bm m_0\times\bm f_c\right)_{\mathcal{A}_{0\perp}} = j\omega_h\left(\bm\varphi_h,\bm f_c\right) - \frac{j\omega_h}{2}\sum_{i,j}\left(\bm\varphi_h,\psi_{ij}\,\bm f_c\right)\,a_ia_j\,,\\
    &\left(\bm\varphi_h, \delta\bm m_\perp\times\bm f_c\right)_{\mathcal{A}_{0\perp}} = -j\omega_h\sum_{i}\left(\bm\varphi_h,\bm m_0\cdot\bm f_c\,\bm\varphi_i\right)\,a_i\,,\\
    \end{aligned}
\end{equation}
while the terms due to $\bm f_d$ are expressed by the following relations:
\begin{equation}\label{eq:proj_nonlin_forc_2}
\begin{aligned}
    &\bm m\times(\bm m\times\bm f_d)\big |_\perp = (1+\delta m_0)\,\bm m_0\cdot \bm f_d\,\delta \bm m_\perp + \delta\bm m_\perp\cdot \bm f_d\,\delta \bm m_\perp + \bm m_0\times(\bm m_0\times\bm f_d)\,,\\
    &\left(\bm\varphi_h, (1+\delta m_0)\,\bm m_0\cdot \bm f_d\,\delta \bm m_\perp\right)_{\mathcal{A}_{0\perp}} = j\omega_h\sum_i\left(\bm m_0\times\bm\varphi_h,\bm m_0\cdot \bm f_d\,\bm\varphi_i\right)\,a_i - \frac{j\omega_h}{2}\sum_{i,j,k}\left(\bm m_0\times\bm\varphi_h,\psi_{ij}\,\bm m_0\cdot \bm f_d\,\bm\varphi_i\right)\,a_ia_ja_k\,,\\
    &\left(\bm\varphi_h, \delta\bm m_\perp\cdot \bm f_d\,\delta \bm m_\perp\right)_{\mathcal{A}_{0\perp}} = j\omega_h\sum_{i,j}\left(\bm m_0 \times\bm\varphi_h,\bm \varphi_j\cdot\bm f_d\,\bm\varphi_i\right)\,a_ia_j\,,\\
    &\left(\bm\varphi_h, \bm m_0\times(\bm m_0\times\bm f_d)\right)_{\mathcal{A}_{0\perp}} = j\omega_h\left(\bm\varphi_h,\bm m_0\times\bm f_d\right)\,.
    \end{aligned}
\end{equation}
Once that the terms due to the driving force of  magnetization dynamics are projected on the normal modes basis, we arrive at the following model:
\begin{equation}\label{eq:nonlin_forc_odes_mode}
\begin{aligned}
(1+\alpha_\mathrm{G}^2)\,\dot{a}_h = &j\omega_h\, b_{h} + j\omega_h\sum_i b_{hi}a_i + j\omega_h\sum_{i,j}c_{hij} a_ia_j\,+ \frac{j\omega_h}{2}\sum_{i,j,k}d_{hijk}a_ia_ja_k\,+\\
&\frac{j\omega_h}{4}\sum_{i,j,k,l}e_{hijkl}a_ia_ja_ka_l\,+ \frac{j\omega_h}{4}\sum_{i,j,k,l,m}f_{hijklm}a_ia_ja_ka_la_m\,,
\end{aligned}
\end{equation}
where:
\begin{equation}\label{eq:coeff_odes_nonlin_forc_mode}
 \begin{aligned}
 &b_h = b_h^f\,,\\
 &b_{hi} = \delta_{hi}+\alpha_\mathrm{G}\,\,b_{hi}^\alpha + b_{hi}^f\,,\\
 &c_{hij} = c_{hij}^0 +\alpha_\mathrm{G}\,c_{hij}^\alpha+c_{hij}^f\,,\\
 & d_{hijk} = d_{hijk}^0+\alpha_\mathrm{G}\,d_{hijk}^\alpha + d_{hijk}^f \,,\\
 & e_{hijkl} = e_{hijkl}^0+\alpha_\mathrm{G}\,e_{hijkl}^\alpha\,,\\
 & f_{hijklm} = \alpha_\mathrm{G}\,f_{hijklm}^\alpha\,,\\
 \end{aligned}
\end{equation}
with $b_h^f, b_{hi}^f,\,c_{hij}^f,\,d_{hijk}^f$ being the coefficients inferred from equation \eqref{eq:proj_nonlin_forc_1} which multiply the product of 0 (purely additive), 1, 2, and 3 normal modes amplitudes, respectively. The above model with formulas for the coefficients is reported in table \ref{tab:NMM} of section Results.

\section*{Acknowledgements}
The financial support of the FWF project I 4917 and the support from the Christian Doppler Laboratory "Advanced Magnetic Sensing and Materials" and the commercial Partner Infineon is acknowledged. We acknowledge financial support from the Horizon 2020 Framework Programme of the European Commission under FET-Open grant agreement no. 899646 (k-NET).

\section*{Author contributions}
M.d'A., D.S. and C.S. planned the study. S.P. and M.d'A. developed the formalism of the normal modes-based model and the numerical simulation code with consultation from D.S and F.B. All authors discussed the results and wrote the manuscript.

\end{document}